%% file: ex_article.tex
\documentclass[review,onefignum,onetabnum]{siamart251216}
\usepackage{soul}
\input{ex_shared}

\ifpdf
\hypersetup{
  pdftitle={Universality in the target arrival statistics of non-conservative search processes},
  pdfauthor={J. Giral-Barajas, S. Linn, and P. C. Bressloff}
}
\fi

\linenumbers

\begin{document}

\maketitle

% REQUIRED
\begin{abstract}
Stochastic search processes in which agents are continuously inserted into and removed from a target search domain are fundamental to a wide class of physical and artificial systems. The theory of such non-conservative search processes is, however, much less developed than for search processes with a fixed number of agents. Recent work on the non-conservative case has characterized the fastest first-passage time (FPT) to a target, after which the search process is terminated. Here we consider the complementary problem of continuous arrival statistics to a target. In particular, we derive the full time-dependent distribution of target arrivals, which we use to show that the inter-arrival time distribution is exactly exponential in steady-state regardless of the underlying search process. Thus, counterintuitively, the arrival statistics of non-conservative search are both universal and much simpler than sequential search-and-capture processes involving a fixed number of agents. Our methods also yield an exact expression for the fastest FPT distribution with time-dependent insertion, thereby generalizing previous work on fixed-rate insertion for non-conservative search. Altogether these results have major implications for target resource accumulation, wherein the delivery of resources is counter-balanced by their downstream consumption.
\end{abstract}

% REQUIRED

% REQUIRED

%%%%%%%%%%%%%%%%%%%%%%%%%%%%%%%%%%%%%%%%%%%%%%%%%%%%%%%%%%%%%%%%%%%%%%%%%%%%%%%%%
%%%%%%%%%%%%%%%%%%%%%%%%%%%%%%%%%%%%%%%%%%%%%%%%%%%%%%%%%%%%%%%%%%%%%%%%%%%%%%%%%
%%%%%%%%%%%%%%%%%%%%%%%%%%%%%%%%%%%%%%%%%%%%%%%%%%%%%%%%%%%%%%%%%%%%%%%%%%%%%%%%%
%%%%%%%%%%%%%%%%%%%%%%%%%%%%%%%%%%%%%%%%%%%%%%%%%%%%%%%%%%%%%%%%%%%%%%%%%%%%%%%%%

\section{Introduction}Stochastic processes in which searching agents are continuously inserted into and removed from a domain are ubiquitous in physical and artificial systems. These dynamics arise, for instance, in the context of biology and population science regarding birth-death processes \cite{Kendall1948, Novozhilov2006, Crawford2012, komarova2025population}, sometimes also with immigration \cite{dessalles2018,Xu_2018} or catastrophes \cite{Brockwell_Gani_Resnick_1982, MANGEL19931, KYRIAKIDIS1994239, KYRIAKIDIS1995346, Renshaw01011997, Kapodistria_Phung-Duc_Resing_2016}. Other terminology used to describe agents that stochastically join and leave a domain include dynamic redundancy and mortality \cite{campos_dynamic_2024, linn_dynamic_2026}, synthesis and degradation \cite{Raj2006, boyer2026}, and entry and exit \cite{Dunne1988, pal2026geo}. While the precise mechanisms governing insertion and removal may seem incomparable between, say, biological and economic systems, the corresponding mathematics is agnostic to the application.

In addition to the relevance of changes in population size, the domains occupied by agents often contain target regions of interest, yielding questions of stochastic search. For example, at the level of the cell nucleus, transcription factors seek specific DNA binding sites to initiate RNA transcription \cite{alberts_molecular_2015, rodriguez_intrinsic_2019}. Transcribed RNA then navigates the cytoplasm searching for ribosomes while subject to degradation \cite{harries_rna_2019, ham_stochastic_2024}. At the cellular scale, naive T cells continuously enter lymph nodes in search of antigen-presenting cells \cite{parkin_overview_2001, currie_stochastic_2012, wong_first_2025}, activating an immune response only upon successful encounters within their finite residence times \cite{li_reliable_2024}.

Stochastic search with a fluctuating--or `non-conservative'--number of agents is a rapidly growing field of interest but analytical developments remain limited compared to its fixed-number counterpart. Existing work has largely adopted an agent-centric perspective, characterizing first-passage properties such as the fastest first-passage time (FPT) to a target, at which point the process terminates \cite{campos_dynamic_2024, linn_dynamic_2026,boyer2026, pal2026geo, tung_first_2025, grebenkov_fastest_2025, meyer_optimal_2025,  grebenkov_birth_2026}. The underlying assumption is that it takes the arrival of only a single agent to trigger some downstream event, the timescale of which is given by the fastest FPT. However, there is a complementary target-centric viewpoint in which the process continues beyond the first arrival. That is, whenever an agent arrives to a target it delivers a resource to that target. The resulting accumulation of resources will then depend on the arrival statistics of agents to the target. Previous work in this area has focused on a fixed number of agents that undergo multiple rounds of search-and-capture, rendering the target arrival statistics determined solely by the FPT distribution of a single agent \cite{bressloff_search-and-capture_2019, bressloff_directional_2021, bressloff_first-passage_2021,bressloff_queueing_2020, bressloff_queuing_2021, giral-barajas_stochastic_2025, giral-barajas_resetting_2025}.

In this work we show that the steady-state inter-arrival time distributions of non-conservative stochastic search (NCSS) processes are exactly exponential, a universal result entirely agnostic to the underlying search process. NCSS arrival statistics are therefore surprisingly much simpler than in traditional search-and-capture. In establishing this result we also produce an exact expression for the fastest FPT distribution with time-dependent insertion, generalizing previous work on NCSS which assumes the insertion rate to be fixed. The paper is organized as follows. In \cref{sec:setting} we rigorously define NCSS and consider the particular scenario of Brownian motion on the half-line as an illustrative example. In \cref{sec:maintheory} we describe two complementary approaches to analyze these search processes more generally: In \cref{sec:pathwise} we present a path-wise framework that uses conditional probability mass functions (PMFs) to track agent insertion and removal, and in \cref{sec:queue1} we utilize a natural mapping between NCSS and queueing theory. Our main results are contained in \cref{sec:imp} where we expound on the theory developed in \cref{sec:maintheory} and illustrate these results in a few concrete examples including diffusive and directed search in multiple domains. In \cref{sec:discussion} we discuss interesting directions of future study. We  tabulate notation and definitions, and provide details of the queueing theoretic results in the appendices.

%%%%%%%%%%%%%%%%%%%%%%%%%%%%%%%%%%%%%%%%%%%%%%%%%%%%%%%%%%%%%%%%%%%%%%%%%%%%%%%%%
%%%%%%%%%%%%%%%%%%%%%%%%%%%%%%%%%%%%%%%%%%%%%%%%%%%%%%%%%%%%%%%%%%%%%%%%%%%%%%%%%
%
%%%%%%%%%%%%%%%%%%%%%%%%%%%%%%%%%%%%%%%%%%%%%%%%%%%%%%%%%%%%%%%%%%%%%%%%%%%%%%%%%
%%%%%%%%%%%%%%%%%%%%%%%%%%%%%%%%%%%%%%%%%%%%%%%%%%%%%%%%%%%%%%%%%%%%%%%%%%%%%%%%%

\section{Probabilistic setup and a simple example}
\label{sec:setting}

Consider a domain $\Omega$ of arbitrary dimension and size that contains a totally absorbing target region $\Omega_0\subset \Omega$. An agent subject to prescribed stochastic dynamics in $\Omega$, starting from $\mathbf{x}_0\in\Omega\setminus\Omega_0$, has an intrinsic random FPT to the target defined by
\begin{equation}\label{eq:FPTDef}
    \mathcal{T}(\mathbf{x}_0):=\inf\{t> 0:\mathbf{X}(t)\in \Omega_0,\,\mathbf{X}(0)=\mathbf{x}_0\},
\end{equation}
where $\mathbf{X}(t)$ denotes the position of an agent at time $t$. For notational brevity we hereafter omit the dependence of $\mathcal{T}(\mathbf{x}_0)$ on $\mathbf{x}_0$ and simply write $\mathcal{T}$. 

\begin{figure}[t!]
\centering
  \subfloat[]{\label{fig:DRM_Queue_a}\includegraphics[width=0.42\textwidth]{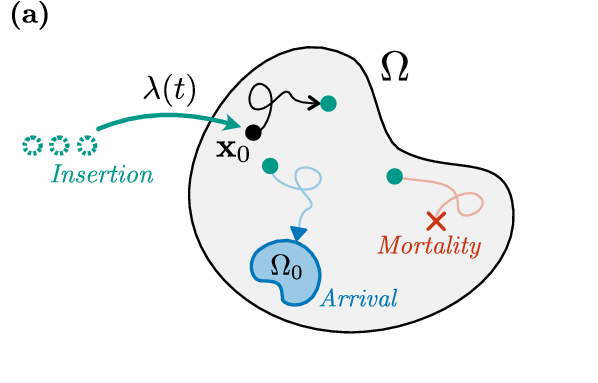}}
  \subfloat[]{\label{fig:DRM_Queue_b}\includegraphics[width=0.42\textwidth]{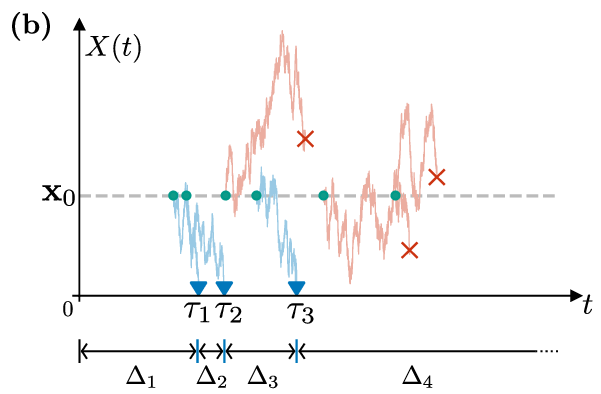}}
  \vspace{-17.5pt}
  \caption{Schematics of non-conservative stochastic search (NCSS). (a) Agents are inserted into the domain $\Omega$ at a fixed position $x_0$ according to a nonhomogeneous Poisson process of rate $\lambda(t)$. Departures correspond either to arrivals to the target $\Omega_0$ (blue triangles) or mortality events (red crosses). (b) Sample paths of a non-conservative 1D Brownian motion search on the half-line ($\Omega_0=\{0\}$) with stochastic insertion and mortality.}
  \label{fig:DRM_Queue}
\end{figure}

We take $\Omega$ to be initially empty with agents independently inserted at $\mathbf{x}_0\in\Omega$ following a nonhomogeneous Poisson process $\{I(t):t\ge0\}$ with rate $\lambda(t)\geq0$. While seeking $\Omega_0$, agents are also independently and irreversibly removed from $\Omega$ after some random lifetime $\mathcal{M}$, which we assume is strictly positive with mean $\mathbb{E}[\mathcal{M}]:=\mu^{-1}\in(0,\infty]$. These agents are said to be `mortal'. Since agents continuously join and leave the search process, the number of agents in $\Omega$ at any given time is random and could even be zero. We illustrate these dynamics in \cref{fig:DRM_Queue_a}.

The probability that a single mortal agent, independent of others, finds the target before dying is
\begin{equation}
    \label{eq:pmu}
    p := \mathbb{P}(\mathcal{T}< \mathcal{M}) = \int_0^\infty S_\mathcal{M}(t)\,\textup{d}F_\mathcal{T}(t),
\end{equation}
where $S_\mathcal{M}(t):=\mathbb{P}(\mathcal{M}>t)$ denotes the survival function of the agent's lifetime (assuming it is inserted at $t=0$), and $F_{\mathcal{T}}(t):= \mathbb{P}(\mathcal{T}\leq t)$ denotes the cumulative distribution function of the FPT of the underlying search process. We assume throughout that $p>0$, which simply means that any agent can find the target before dying. In \cref{app:p_proof} we show that this assumption holds under very mild conditions on the search process and provide concrete examples fulfilling these conditions.

Finally we denote by $\tau_n$ the time of the $n$th arrival to the target. The $n$th target inter-arrival time is therefore $\Delta_n:=\tau_n-\tau_{n-1}$ with $\tau_0\equiv 0$ (see \cref{fig:DRM_Queue_b}). If $\{\Delta_n\}_{n=1}^\infty$ is a sequence of independent and identically distributed (iid) random variables, we denote by $\Delta$ the common inter-arrival time random variable.

Before demonstrating the theory in full, we first showcase a simple example of NCSS for Brownian motion on the half line and determine the mean arrival rate of agents to the origin using a macroscopic partial differential equation (PDE) framework. Let $\rho(x,t)$ denote the density of agents at position $x$ and time $t$. We take all agents to have fixed insertion rate $\lambda>0$ at $x_0>0$, diffusivity $D>0$, and removal rate $\mu>0$. In particular, these choices mean that the insertion process $I(t)$ is homogeneous Poisson and the lifetime distribution $\mathcal{M}$ is exponential. The PDE governing the evolution of the density is therefore
\begin{align} \label{eq:PDE}
    \partial_t \rho(x,t) = D\partial_{xx} \rho(x,t) - \mu\rho(x,t) + \lambda\delta(x-x_0),
\end{align}
with initial condition $\rho(x,0) = \delta(x-x_0)$ and boundary condition $\rho(0,t)=0$.

In Laplace space, where we denote the Laplace transform of $\rho(x,t)$ by $\tilde{\rho}(x,s)$, \cref{eq:PDE} becomes
\begin{align} \label{eq:LPDE}
    \partial_{xx} \tilde{\rho}(x,s) = \frac{\mu+s}{D}\tilde{\rho}(x,s),
\end{align}
with boundary condition $\tilde{\rho}(0,s)=0$. Solving \cref{eq:LPDE} and imposing continuity of flux at $x=x_0$ yields
\begin{align}
    \tilde{\rho}(x,s) = \begin{cases}
        \alpha(x_0) \frac{\sinh kx}{\sinh kx_0} & 0<x\leq x_0,\\
        \alpha(x_0) e^{-k(x-x_0)} & x_0<x,
    \end{cases}
\end{align}
where $k^2:= (\mu+s)/D$ and
\begin{align}
    \alpha(x_0) := \frac{1+(\lambda/s)}{kD(1+\coth kx_0)}.
\end{align}
Finally we can identify the mean arrival rate as the steady-state flux into the origin,
\begin{align} \label{eq:flux}
    \lim_{t\to\infty} D\partial_x \rho(x,t)\big|_{x=0} = \lim_{s\to 0} sD\partial_x \tilde{\rho}(x,s)\big|_{s=0} = \lambda e^{-\sqrt{\mu x_0^2/D}}.
\end{align}
The right-most expression in \cref{eq:flux} is also the insertion rate $\lambda$ multiplied by the probability that a single mortal agent finds the target before dying as defined in \cref{eq:pmu}. One can moreover recognize this expression as the insertion rate multiplied by the $\mu$-Laplace transform of the FPT density of a single immortal agent with instantaneous resetting at rate $\mu$ to the insertion position $x_0$ \cite{Evans2020}. 

A few comments are in order. First, this PDE approach yields the mean target arrival rate but does not provide information on the corresponding distribution. It moreover requires fixing \emph{a priori} the process by which agents navigate the domain. In the following section we use alternative methods, which are agnostic to agent motility, in order to derive the full target inter-arrival distribution.

%%%%%%%%%%%%%%%%%%%%%%%%%%%%%%%%%%%%%%%%%%%%%%%%%%%%%%%%%%%%%%%%%%%%%%%%%%%%%%%%%
%%%%%%%%%%%%%%%%%%%%%%%%%%%%%%%%%%%%%%%%%%%%%%%%%%%%%%%%%%%%%%%%%%%%%%%%%%%%%%%%%
%
%%%%%%%%%%%%%%%%%%%%%%%%%%%%%%%%%%%%%%%%%%%%%%%%%%%%%%%%%%%%%%%%%%%%%%%%%%%%%%%%%
%%%%%%%%%%%%%%%%%%%%%%%%%%%%%%%%%%%%%%%%%%%%%%%%%%%%%%%%%%%%%%%%%%%%%%%%%%%%%%%%%
\section{Full distribution of the target arrival process} \label{sec:maintheory}
In this section we present two frameworks to derive the full distribution of target arrivals for general NCSS. We first describe a path-wise approach that utilizes various conditional PMFs for the insertion and departure processes to formulate an explicit expression for the target arrival distribution function. We then detail an approach which reformulates non-conservative stochastic search as a queueing process and exploit well-established results in queueing theory to conclude that NCSS target arrival processes are exactly nonhomogeneous Poisson.

\subsection{Path-wise framework}
\label{sec:pathwise}
\begin{figure}[t!]
  \centering
  \includegraphics[width=0.55\textwidth]{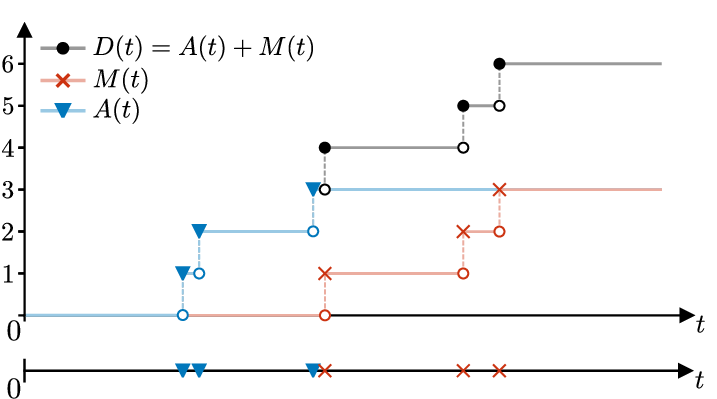}
  \caption{Sample paths of departure, arrival, and mortality processes. Abstracting away from the specific search process, departures (black circles), which consist of both target arrivals (blue triangles) and mortality events (red crosses), can be characterized merely as a counting process in time.}
  \label{fig:Fig2}
\end{figure}

Recall that agents are inserted into the domain $\Omega$ at position $\mathbf{x}_0$ according to a nonhomogeneous Poisson insertion process $\{I(t):t\geq0\}$ with intensity function $\lambda(t)$. We denote the mean insertion function by
\begin{align} \label{eq:It}
    \mathbb{I}(t):=\mathbb{E}[I(t)]=\int_0^t\lambda(t')\,\textup{d}t'.
\end{align}
Now consider the departure process $\{D(t):t\geq0\}$, which counts the number of agents to have left $\Omega$ by time $t$ either via mortality or via contact with $\Omega_0$ (see \cref{fig:Fig2}). Conditioning on the insertion process at time $t$ and using the law of total probability, we can write the probability of $m\geq 0$ departures in $(0,t]$ as
\begin{equation}
    \label{eq:departureProb}
    \mathbb{P}(D(t)=m)=\sum_{n=m}^\infty\mathbb{P}(D(t)=m\mid I(t)=n)\mathbb{P}(I(t)=n).
\end{equation}
Note that the sum in \cref{eq:departureProb} starts from $m$ since there cannot be more departures than insertions at any given time, \textit{i.e.}~$\mathbb{P}(D(t)=m\mid I(t)=n)=0$ for $m>n$.

Having defined the insertion and departure processes we now consider the target arrival process $\{A(t):t\geq0\}$, which counts the number of agents to have reached $\Omega_0$ by time $t$. The PMF of target arrivals in $(0,t]$ can be written as
\begin{equation}
    \label{eq:ArrivalProbTemp}
    \mathbb{P}(A(t)=k)=\sum_{m=k}^\infty\mathbb{P}(A(t)=k\mid D(t)=m)\mathbb{P}(D(t)=m).
\end{equation}
Similar to \cref{eq:departureProb}, the sum in \cref{eq:ArrivalProbTemp} starts from $k$ since there can be no more arrivals to the target than total departures from the domain. Substituting \cref{eq:departureProb} into \cref{eq:ArrivalProbTemp} and re-indexing the sums with $j=m-k$ and $i=n-m$, we obtain
\begin{align}\label{eq:ArrivalProb}
\begin{split}
    \mathbb{P}(A(t)=k)&=\sum_{j=0}^\infty \, \mathbb{P}(A(t)=k\mid D(t)=j+k)\\
    &\qquad\times \sum_{i=0}^\infty\mathbb{P}(D(t)=j+k\mid I(t)=i+j+k)\mathbb{P}(I(t)=i+j+k).
    \end{split}
\end{align}

To make sense of the target arrival PMF, we determine explicit expressions for the three probabilities on the right-hand side of \cref{eq:ArrivalProb}. By definition of the insertion process, we have that
\begin{equation}
    \label{eq:TempProb1}
    \mathbb{P}(I(t)=i+j+k)=\frac{1}{(i+j+k)!}[\mathbb{I}(t)]^{i+j+k} e^{-\mathbb{I}(t)}.
\end{equation}

Next we compute the conditional departure PMF. Conditioning on the insertion process at time $t$, the unordered insertion times are iid with distribution function $F(t')=\mathbb{I}(t')/\mathbb{I}(t)$ for $t'\in(0,t]$. It follows that the probability that an agent is still in $\Omega$ at time $t$ is precisely
\begin{equation}
    \label{eq:Pt}
    q_t:= \frac{1}{\mathbb{I}(t)}\int_0^t\lambda(t')S_\mathcal{S}(t-t')\,\textup{d}t',
\end{equation}
where $S_\mathcal{S}(t):=\mathbb{P}(\mathcal{S}>t)$ is the survival probability of $\mathcal{S}:=\min\{\mathcal{T},\mathcal{M}\}$. Consequently, the number of departures by time $t$ conditioned on the number of insertions is binomial distributed with `success probability' given by $1-q_t$,
\begin{equation}
    \label{eq:TempProb3}
    \mathbb{P}(D(t)=j+k\mid I(t)=i+j+k)=\binom{i+j+k}{i}\, q_t^{\,i}\,(1-q_t)^{j+k}.
\end{equation}

Finally we determine the conditional target arrival PMF, which requires us to distinguish between departures due to mortality and due to target encounters. We define a defective cumulative distribution function which describes the probability that an individual agent finds the target before dying by time $t$,
\begin{align} \label{eq:Find}
    F_{\text{ind}}(t):= \mathbb{P}(\mathcal{T}<\mathcal{M},\,\mathcal{S}<t) = \int_0^t S_\mathcal{M}(t')\,\textup{d}F_{\mathcal{T}}(t').
\end{align}
Comparison with equation \cref{eq:pmu} implies that
\begin{equation}
p=\lim_{t\rightarrow \infty} F_{\text{ind}}(t).
\end{equation}
The probability of a target encounter in $(0,t]$ rather than a mortality event can then be described by the ratio of the expected number of target arrivals to the expected number of all departures,
\begin{equation}
    \label{eq:peff}
    q_t^A := \frac{\int_0^t\lambda(t')\,F_{\text{ind}}(t-t')\,\textup{d}t'}{\int_0^t\lambda(t')\,F_{\mathcal{S}}(t-t')\,\textup{d}t'}.
\end{equation}
with
\begin{align} \label{eq:FS}
    F_\mathcal{S}(t) := \mathbb{P}(\mathcal{S}\leq t) = 1 - S_\mathcal{T}(t)S_\mathcal{M}(t).
\end{align}

Since agents evolve independently, departure times are therefore iid draws from the distribution in \cref{eq:peff}, and each departure is independently classified as an arrival to target with probability $q_t^A$, hence
\begin{equation}
    \label{eq:TempProb2}
    \mathbb{P}(A(t)=k\mid D(t)=j+k)=\binom{j+k}{k}(q_t^A)^k(1-q_t^A)^j.
\end{equation}

Substituting \cref{eq:TempProb1,eq:TempProb3,eq:TempProb2} into \cref{eq:ArrivalProb} and simplifying, we obtain
\begin{align}
    \mathbb{P}(A(t)=k) = \frac{1}{k!} \bigl[q_t^A(1-q_t)\mathbb{I}(t)\bigr]^k e^{-q_t^A(1-q_t)\mathbb{I}(t)},
\end{align}
hence $A(t)$ is a Poisson random variable with mean
\begin{equation} \label{eq:ayy}
    \mathbb{A}(t) := \mathbb{E}[A(t)] = q_t^A\,(1-q_t)\,\mathbb{I}(t).
\end{equation}
Further substitution and simplification ultimately yields
\begin{equation}
    \label{eq:ArrMeanFuncDerived}
    \mathbb{A}(t) = \int_0^t \lambda(t-t')F_{\text{ind}}(t') \,\textup{d}t'.
\end{equation}

Altogether in this section we have shown that the number of target arrivals for a fixed time $t>0$ is Poisson distributed with mean given by \cref{eq:ArrMeanFuncDerived}. To conclude that the target arrival process is in fact exactly nonhomogeneous Poisson, it remains to show that $A(t)$ has independent increments.\footnote{Not all counting processes with a marginal Poisson distribution have independent increments. One example is a Cox process, also called a doubly stochastic Poisson process, which is a Poisson process whose rate function itself evolves stochastically.} Though possible, establishing this additional criterion is considerably involved and cumbersome. Alternatively, in the following subsection we exploit a mapping between NCSS and queueing theory to conclude that the law of the arrival process is indeed the law of a non-homogeneous Poisson process.

%%%%%%%%%%%%%%%%%%%%%%%%%%%%%%%%%%%%%%%%%%%%%%%%%%%%%%%%%%%%%%%%%%%%%%%%%%%%%%%%%
%%%%%%%%%%%%%%%%%%%%%%%%%%%%%%%%%%%%%%%%%%%%%%%%%%%%%%%%%%%%%%%%%%%%%%%%%%%%%%%%%
%
%%%%%%%%%%%%%%%%%%%%%%%%%%%%%%%%%%%%%%%%%%%%%%%%%%%%%%%%%%%%%%%%%%%%%%%%%%%%%%%%%
%%%%%%%%%%%%%%%%%%%%%%%%%%%%%%%%%%%%%%%%%%%%%%%%%%%%%%%%%%%%%%%%%%%%%%%%%%%%%%%%%
%%%%%%%%%%%%%%%%%%%%%%%%%%%%%%%%%%%%%%%%%%%%%%%%%%%%%%%%%%%%%%%%%%%%%%%%%%%%%%%%%
%%%%%%%%%%%%%%%%%%%%%%%%%%%%%%%%%%%%%%%%%%%%%%%%%%%%%%%%%%%%%%%%%%%%%%%%%%%%%%%%%
%
%%%%%%%%%%%%%%%%%%%%%%%%%%%%%%%%%%%%%%%%%%%%%%%%%%%%%%%%%%%%%%%%%%%%%%%%%%%%%%%%%
%%%%%%%%%%%%%%%%%%%%%%%%%%%%%%%%%%%%%%%%%%%%%%%%%%%%%%%%%%%%%%%%%%%%%%%%%%%%%%%%%
\subsection{Queueing theoretic framework}\label{sec:queue1}
Queueing theory is a mathematical framework used to study the insertion and processing of agents in a queue. Queues are canonically and uniquely described using Kendall’s shorthand notation $A/B/c$, where $A$ describes the inter-insertion time distribution, $B$ the service time distribution, and $c\in\{1,2,\dots\}$ the number of servers \cite{kendall_stochastic_1953}. Also standard is the use of $M_t$ to denote a nonhomogeneous Poisson insertion process, $M$ a homogeneous Poisson insertion process, and $G$ a general service-time distribution \cite{eick_physics_1993,cooper_introduction_1981}. We provide a brief introduction to fundamentals of queues with Poisson insertions in \cref{app:queues}.

\begin{figure}[htbp]
  \subfloat[]{\label{fig:Fig3_a}\includegraphics[width=0.4841\textwidth]{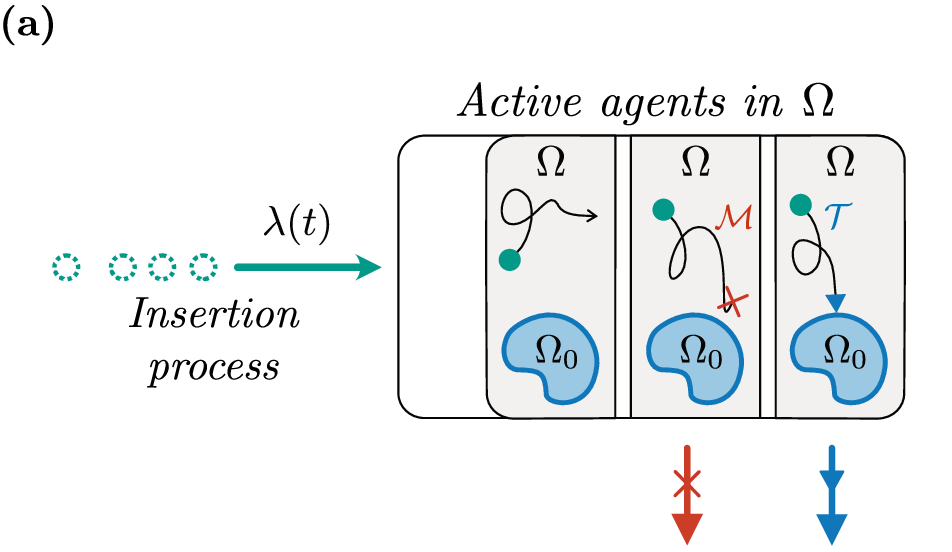}}
  \subfloat[]{\label{fig:Fig3_b}\includegraphics[width=0.5159\textwidth]{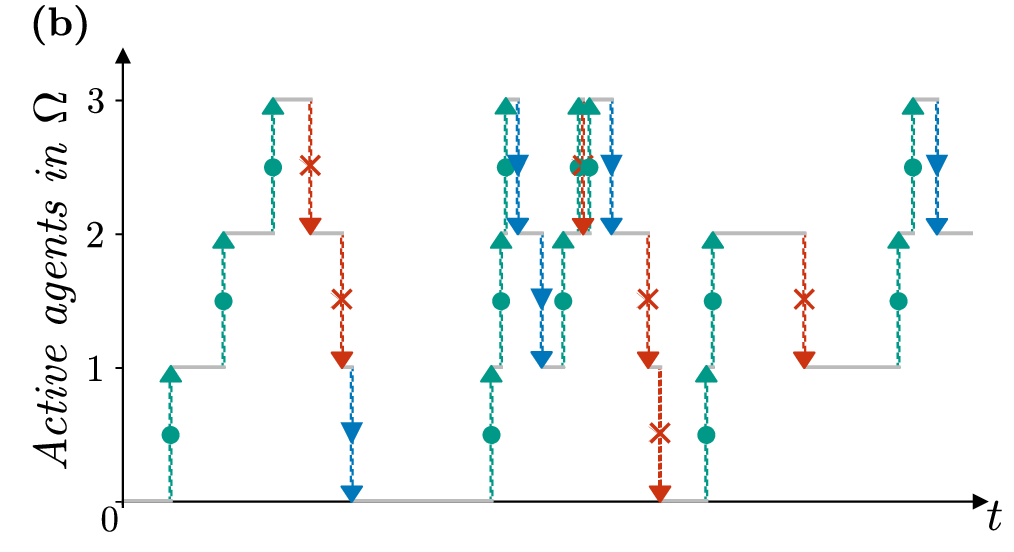}}
  \vspace{-17.5pt}
  \caption{Mapping of NCSS to a queueing process. (a) Searchers are continuously inserted into the queue $\Omega$ according to a nonhomogeneous Poisson process of rate $\lambda(t)$. They remain there for a random time $\mathcal{S}=\min\{\mathcal{T},\mathcal{M}\}$ before departing the system. (b) Sample path of the number of agents in $\Omega$ modeled as a queueing system.}
  \label{fig:Fig3}
\end{figure}

In an NCSS process, the domain $\Omega$ is the queue and agents in $\Omega$ are customers in the queue (see \cref{fig:Fig3}). Customer insertions are therefore governed by $I(t)$ and service times are described by $\mathcal{S}:=\min\{\mathcal{T},\mathcal{M}\}$ with general (non-Markovian) distribution function given by \cref{eq:FS}. Since every agent searches independently of others--therefore independently of the number of others--in $\Omega$, the queue is said to have an infinite number of servers. Altogether we have that NCSS is precisely described by an $M_t/G/\infty$ queue.

A well-established result concerning $M_t/G/\infty$ queues (see also \cref{app:queues}) is that the departure process $D(t)$ is nonhomogeneous Poisson \cite{foley1982non} with mean function
\begin{equation}
\label{eq:Dtmeanfunc}
    \mathbb{D}(t):=\mathbb{E}[D(t)]=\int_0^t \lambda(t-t')F_\mathcal{S}(t') \,\textup{d}t'.
\end{equation}
In light of this result, we can decompose the departure process into target arrivals and mortality events. Owing to the independence of mortality and target arrival times, the service-time density (if it exists) is given by
\begin{equation}
\label{eq:fs}
    \frac{d}{dt}F_\mathcal{S}(t) = S_\mathcal{M}(t) \mathbb{P}(\mathcal{T}=t) + S_\mathcal{T}(t) \mathbb{P}(\mathcal{M}=t),
\end{equation}
where the first and second terms on the right-hand side of \cref{eq:fs} describe the rates at which agents contribute to target arrivals and mortalities, respectively, at time $t$. Combining \cref{eq:Dtmeanfunc,eq:fs} yields $\mathbb{D}(t) = \mathbb{A}(t)+\mathbb{M}(t)$ where $\mathbb{A}(t)$ and $\mathbb{M}(t)$ are the mean functions for target arrivals and mortalities, respectively, given by
\begin{subequations}
    \begin{eqnarray}
        \mathbb{A}(t) &:=&\int_0^t \lambda(t-t')\int_0^{t'}S_\mathcal{M}(t'')\,\textup{d}F_\mathcal{T}(t'')  \,\textup{d}t', \label{eq:ArrMeanFunc}\\
        \mathbb{M}(t) &:=&\int_0^t  \lambda(t-t')\int_0^{t'}S_\mathcal{T}(t'')\,\textup{d}F_\mathcal{M}(t'') \,\textup{d}t'. \label{eq:MortMeanFunc}
    \end{eqnarray}
\end{subequations}
Note that substituting \cref{eq:Find} into \cref{eq:ArrMeanFunc} recovers the previous result \cref{eq:ArrMeanFuncDerived} for the mean number of target arrivals.

Since the departure process is nonhomogeneous Poisson and each agent independently contributes to the arrival and mortality counts, thinning properties of Poisson processes\footnote{The thinning property states that randomly removing events from a Poisson process creates two or more independent Poisson processes. If an original process has a rate of $\lambda$, and each event is kept with probability $p$, then the kept events form a new Poisson process with rate $\lambda p$.} give the following result: the full target arrival and mortality processes, $A(t)$ and $M(t)$, are independent Poisson processes with mean functions $\mathbb{A}(t)$ and $\mathbb{M}(t)$ given by equations \cref{eq:ArrMeanFunc} and \cref{eq:MortMeanFunc}, respectively. Hence,
the map between NCSS and an $M_t/G/\infty$ queue implies that the arrival of agents to a target is distributed exactly according to a nonhomogeneous Poisson process. Moreover, the dynamics of target arrivals is completely described by the insertion rate and single-agent dynamics.

%%%%%%%%%%%%%%%%%%%%%%%%%%%%%%%%%%%%%%%%%%%%%%%%%%%%%%%%%%%%%%%%%%%%%%%%%%%%%%%%%
%%%%%%%%%%%%%%%%%%%%%%%%%%%%%%%%%%%%%%%%%%%%%%%%%%%%%%%%%%%%%%%%%%%%%%%%%%%%%%%%%
%
%%%%%%%%%%%%%%%%%%%%%%%%%%%%%%%%%%%%%%%%%%%%%%%%%%%%%%%%%%%%%%%%%%%%%%%%%%%%%%%%%
%%%%%%%%%%%%%%%%%%%%%%%%%%%%%%%%%%%%%%%%%%%%%%%%%%%%%%%%%%%%%%%%%%%%%%%%%%%%%%%%%
\section{Implications of a nonhomogeneous Poisson target arrival process}\label{sec:imp}
\begin{figure}[htbp]
    \centering
    \subfloat[]{\label{fig:Fig4_a}\includegraphics[height=2.44cm]{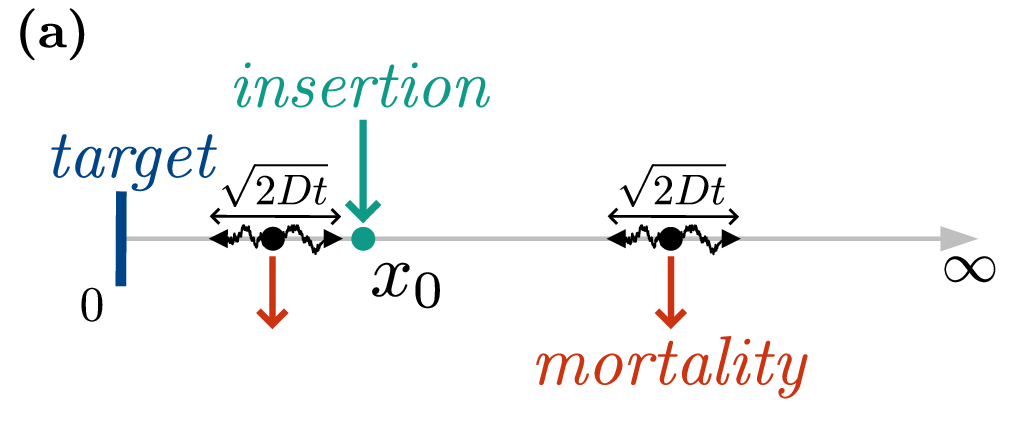}}
    \subfloat[]{\label{fig:Fig4_b}\includegraphics[height=2.44cm]{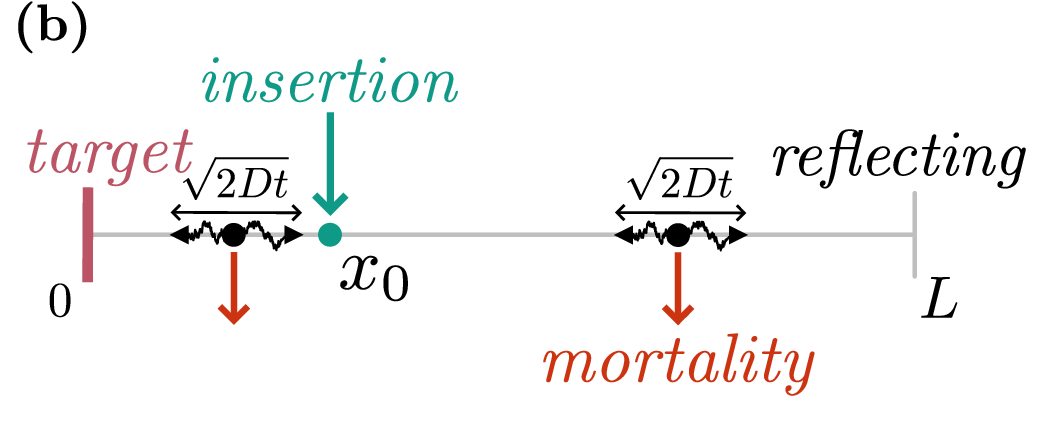}}\\
    \vspace{-17.5pt}
    \subfloat[]{\label{fig:Fig4_c}\includegraphics[height=2.44cm]{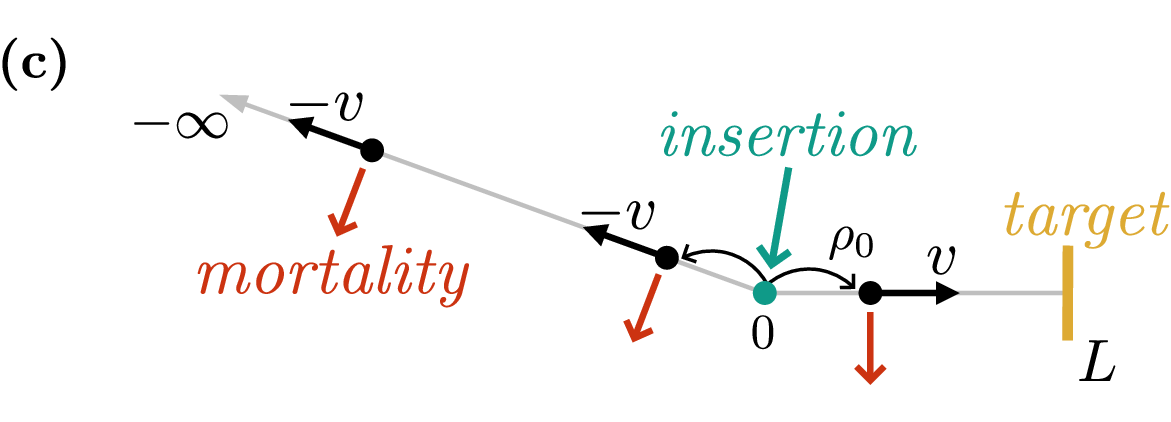}}
    \vspace{-17.5pt}
    \caption{Example scenarios for diffusive and directed NCSS. (a) Agents diffusing on the half-line with a target at $x=0$. Agents are inserted at $x_0>0$ and have exponentially distributed mortality times. (b) Agents diffusing in a finite interval $[0,L]$ with a target at $x=0$ and a reflecting boundary at $x=L$. Agents are inserted at $x_0$ and have uniformly distributed mortality times. (c) Agents enter either the right branch with probability $\rho_0$ or the left branch with probability $1-\rho_0$. Agents move at constant speed $v$ away from the node until either dying or finding the target at $x=L$. Mortality times are exponentially distributed.}
\label{fig:Fig4}
\end{figure}

The universal nature of the arrival process derived in \cref{sec:maintheory} is initially surprising; our limited assumptions on the system allow for agents to navigate their environment in any number of ways including diffusive and directed search. The environment itself can also have complicated features including obstacles and non-trivial boundary interactions. The critical observation is that the controlled continuous insertion of agents ultimately drives the timescale of target encounters, and mortality simply serves as a thinning of the number of agents who ultimately find the target. We will explore various important consequences of these results using three examples:

\medskip
\begin{enumerate}[(a)]
    \item Brownian motion on the positive half line with insertion position $x_0>0$, diffusivity $D>0$, exponentially distributed mortality times $\mathcal{M}\sim\textup{Exp}(\mu)$, and a target at the origin, see \cref{fig:Fig4_a}.\label{ex:a}
    
    \item Brownian motion in the interval $[0,L]$ with insertion position $x_0\in(0,L)$, diffusivity $D>0$, uniformly distributed mortality times $\mathcal{M}\sim\textup{U}(0,2/\mu)$, reflecting boundary at $L$, and a target at the origin, see \cref{fig:Fig4_b}.\label{ex:b}
    
    \item Directed search with two branches $(-\infty,0]\cup[0,L]$ and insertion at $x_0=0$; upon insertion agents move with probability $\rho_0$ along the right branch with velocity $v>0$, and  with probability $1-\rho_0$ along left branch with velocity $-v<0$, each with exponentially distributed mortality times $\mathcal{M}\sim\textup{Exp}(\mu)$, see \cref{fig:Fig4_c}.\label{ex:c}
\end{enumerate}

\medskip
\noindent Hereafter, whenever a distinction is needed between the computed quantities for these examples we use the corresponding letters (a, b, or c) as identifying indices. In each case, we can derive exact expressions for
the FPT distribution of a single immortal agent inserted at time $t=0$,
\begin{subequations}
\begin{eqnarray}
    F_{\mathcal{T}\text{a}}(t) & = & \textup{erfc}\left(\sqrt{\frac{x_0^2}{4Dt}}\right),\label{eq:FT1}\\
    F_{\mathcal{T}\text{b}}(t) & = & 1 - \frac{4}{\pi}\sum_{n=1}^\infty \frac{1}{2n-1}\,\sin\Big[\frac{(2n-1)\pi x_0}{2L}\Big]\,\textup{exp}\Big[\frac{-(2n-1)^2\pi^2Dt}{4L^2}\Big],\label{eq:FT2}\\
    \qquad F_{\mathcal{T}\text{c}}(t) & = &\rho_0\mathbbm{1}_{\{t\geq L/v\}}. \label{eq:FT3}
\end{eqnarray}
\end{subequations}
\Cref{eq:FT1,eq:FT2} can be obtained following standard methods \cite[Chapters~2~\&~3]{redner_guide_2001}. The FPT associated with \cref{eq:FT3} can be defined directly,
\begin{equation}\label{eq:Tc}
    \mathcal{T}\text{c} = \begin{cases}
        L/v\quad&\mbox{with probability }\rho_0,\\
        \infty\quad&\mbox{with probability }1-\rho_0.
    \end{cases}
\end{equation}
The law of total probability therefore indicates that $ F_{\mathcal{T}\text{c}}(t) = \rho_0\mathbbm{1}_{\{t\geq L/v\}}$.

The success probabilities for examples \labelcref{ex:a,ex:b,ex:c} are obtained by direct substitution of \cref{eq:FT1,eq:FT2,eq:FT3} into \cref{eq:pmu}:
\begin{subequations}\label{eq:exact_p}
\begin{eqnarray}
    p_{\text{a}} & = & e^{-\sqrt{\mu x_0^2/D}},\label{eq:exact_p1}\\
    p_{\text{b}} & = & 1-\frac{8\mu L^2}{D\pi^3}\sum_{n=1}^\infty\frac{1}{(2n-1)^3}\,\sin\left[\frac{(2n-1)\pi x_0}{2L}\right]\exp\left[\frac{-(2n-1)^2\pi^2D}{2\mu L^2}\right],\label{eq:exact_p2}\\
    \qquad p_{\text{c}} & = & \rho_0 e^{-\mu L/v}\label{eq:exact_p3}.
\end{eqnarray}
\end{subequations}
Note that $\lambda  p_{\text{a}}$ is in agreement with the steady-state flux calculated in equation \cref{eq:flux}.

Finally, substituting the FPT distributions in \cref{eq:FT1,eq:FT2,eq:FT3} and the corresponding mortality time distributions into \cref{eq:Find} then determines $F_{\textup{ind}}(t)$. In \labelcref{ex:a}, the integral for $F_{\textup{ind}}(t)$ cannot be computed explicitly and is therefore approximated via quadrature. In \labelcref{ex:b}, the eigenfunction expansion of $F_{\mathcal{T}\text{b}}(t)$ in \cref{eq:FT2} can be used to find $F_{\textup{ind}}^\text{b}(t)$. The resulting expression is
\begin{equation}\label{eq:Find2}
    F_{\textup{ind}}^\text{b}(t) = \begin{cases}
        \Upsilon_\text{b}(t)\quad&0<t<2/\mu,\\
        p_\text{b}\quad&t\geq2/\mu,
    \end{cases}
\end{equation}
where
\small
\begin{equation}\label{eq:Find2_aux}
    \Upsilon_\text{b}(t) = 1-\left(1-\frac{\mu t}{2}\right)\frac{2}{L}\sum_{n=1}^\infty\frac{\sin(k_nx_0)}{k_n}e^{-Dk_n^2t}-\frac{\mu}{LD}\sum_{n=1}^\infty\frac{\sin(k_nx_0)}{k_n^3}\left(1-e^{-Dk_n^2}\right),
\end{equation}
\normalsize
with $k_n:=(2n-1)\pi/2L$. Finally, in \labelcref{ex:c}, it is straightforward to show that
\begin{equation}\label{eq:Find3}
    F_{\textup{ind}}^\text{c}(t) = \rho_0e^{-\mu\frac{L}{v}}\mathbbm{1}_{\{t>L/v\}} = p_{\text{c}}\mathbbm{1}_{\{t>L/v\}}.
\end{equation}

%%%%%%%%%%%%%%%%%%%%%%%%%%%%%%%%%%%%%%%%%%%%%%%%%%%%%%%%%%%%%%%%%%%%%%%%%%%%%%%%%
%%%%%%%%%%%%%%%%%%%%%%%%%%%%%%%%%%%%%%%%%%%%%%%%%%%%%%%%%%%%%%%%%%%%%%%%%%%%%%%%%
%
%%%%%%%%%%%%%%%%%%%%%%%%%%%%%%%%%%%%%%%%%%%%%%%%%%%%%%%%%%%%%%%%%%%%%%%%%%%%%%%%%
%%%%%%%%%%%%%%%%%%%%%%%%%%%%%%%%%%%%%%%%%%%%%%%%%%%%%%%%%%%%%%%%%%%%%%%%%%%%%%%%%
\subsection{Fastest first-passage time distribution}\label{subsec:2ffpt_dist} A quantity often of interest in stochastic search problems is the random time taken for the target to be found for the first time by any agent \cite{redner_guide_2001}. Given the mean target arrival function in \cref{eq:ArrMeanFunc}, the probability that the target is not found by any agent by time $t$ is immediate. In particular, this quantity corresponds to the survival probability of the first target inter-arrival time, which is given by
\begin{equation}
    \label{eq:SurvivalPoisson}
    S_{\Delta_1}(t) := \mathbb{P}(\Delta_1>t) =e^{-\mathbb{A}(t)}.
\end{equation}
The expression in \cref{eq:SurvivalPoisson} extends previous results \cite{linn_dynamic_2026, boyer2026} from fixed-rate to time-dependent insertion.

\begin{figure}[b!]
  \centering
  \includegraphics[width=0.6\textwidth]{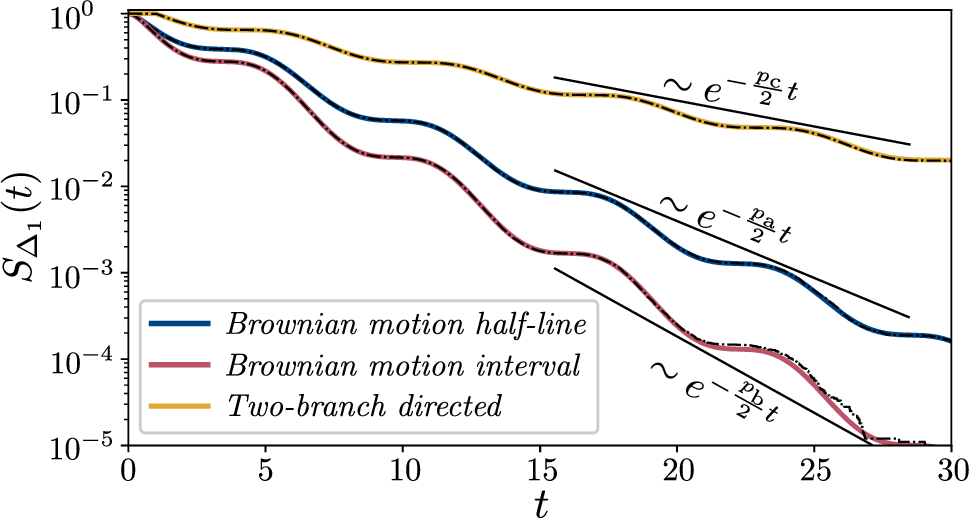}
  \caption{NCSS FPT survival probability $S_{\Delta_1}(t)$ with time-periodic oscillatory insertion rate $\lambda(t)=(1+\cos t)/2$. Solid curves denote the analytical distribution as in \cref{eq:SurvivalPoisson} and dashed curves denote distributions constructed from simulations. Discrepancies between exact and simulated results are due to the rarity of sampling FPTs at large times.}
  \label{fig:Fig5}
\end{figure}

In \cref{fig:Fig5} we illustrate the influence of time-dependent insertion on the FPT distribution where we set $\lambda(t) = (1+\cos t)/2$, hence the insertion rate periodically oscillates between zero and one. Computing the distribution \cref{eq:SurvivalPoisson} for each example in \cref{fig:Fig4} involves varied approaches: In example \labelcref{ex:a} one must approximate the defining integral in \cref{eq:ArrMeanFuncDerived} via quadrature prior to substitution into \cref{eq:SurvivalPoisson}. In \labelcref{ex:b}, the integral expression for $\mathbb{A}_b(t)$ can be simplified using the angle-addition identity of $\cos t$\footnote{$\cos(t-t')=\cos t\cos t'+\sin t\sin t'$}, yielding
\begin{equation}\label{eq:Arr2}
    \mathbb{A}_\text{b}(t) = \frac{1}{2}\left[I_1(t)+\cos(t)I_2(t)+\sin(t)I_3(t)\right],
\end{equation}
with
\begin{subequations}\label{eq:Arr2_aux}
    \begin{eqnarray}
        I_1(t) & = & \int_0^tF_{\textup{ind}}^\text{b}(t')\,\textup{d}t',\label{eq:Arr2_aux1}\\
        I_2(t) & = & \int_0^t\cos(t')F_{\textup{ind}}^\text{b}(t')\,\textup{d}t',\label{eq:Arr2_aux2}\\
        I_3(t) & = & \int_0^t\sin(t')F_{\textup{ind}}^\text{b}(t')\,\textup{d}t'.\label{eq:Arr2_aux3}
    \end{eqnarray}
\end{subequations}
After this simplification, \cref{eq:Arr2_aux1,eq:Arr2_aux2,eq:Arr2_aux3} are approximated via quadrature. In \labelcref{ex:c}, $\mathbb{A}_\text{c}(t)$ can be computed directly,
\begin{equation}\label{eq:Arr3}
    \mathbb{A}_\text{c}(t) = \frac{\rho_0}{2v}e^{-\frac{L\mu}{v}}\left[v\sin(t-L/v)+vt-L\right]\mathbbm{1}_{\{t>L/v\}}.
\end{equation}
While the oscillatory nature of the insertion process evidently induces oscillations in the survival probability, the non-oscillatory component of the insertion rate function determines the persistent rate of exponential decay. 

%%%%%%%%%%%%%%%%%%%%%%%%%%%%%%%%%%%%%%%%%%%%%%%%%%%%%%%%%%%%%%%%%%%%%%%%%%%%%%%%%
%%%%%%%%%%%%%%%%%%%%%%%%%%%%%%%%%%%%%%%%%%%%%%%%%%%%%%%%%%%%%%%%%%%%%%%%%%%%%%%%%
%
%%%%%%%%%%%%%%%%%%%%%%%%%%%%%%%%%%%%%%%%%%%%%%%%%%%%%%%%%%%%%%%%%%%%%%%%%%%%%%%%%
%%%%%%%%%%%%%%%%%%%%%%%%%%%%%%%%%%%%%%%%%%%%%%%%%%%%%%%%%%%%%%%%%%%%%%%%%%%%%%%%%
\subsection{Markovianity emerges in steady-state}\label{subsec:3emerge}
\Cref{eq:ArrMeanFuncDerived} provides a complete description of the mean number of target arrivals in the most general case of NCSS in which the insertion rate may be time-dependent. By fixing the insertion rate, $\lambda(t)\equiv\lambda$, the number of agents in $\Omega$ simplifies to an $M/G/\infty$ queue (see \cref{app:queues}) and we find the time-dependent probability flux into the target to be
\begin{equation}
    \label{eq:IntensityFuncLambda}
    a_\lambda(t) := \frac{\textup{d}}{\textup{d}t} \mathbb{A}(t)\big|_{\lambda(t)=\lambda} = \lambda F_{\text{ind}}(t).
\end{equation}
It follows that the steady-state flux into the target is constant,
\begin{align} \label{eq:stat}
    \lim_{t\to\infty} a_\lambda(t) = \lambda p.
\end{align}

\begin{figure}[htbp]
  \centering
  \subfloat[]{\label{fig:Fig6_a}\includegraphics[height=3.7cm]{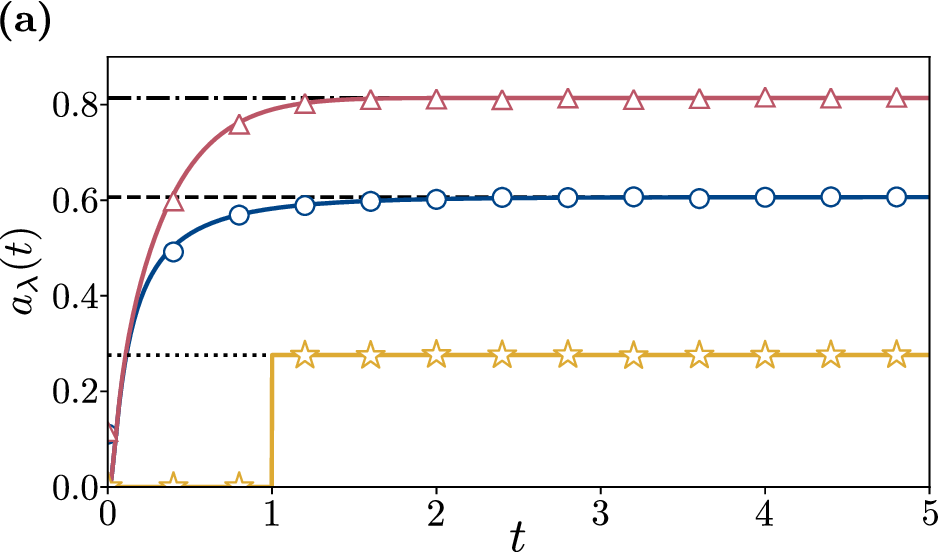}}\hspace{0.065cm}
  \subfloat[]{\label{fig:Fig6_b}\includegraphics[height=3.7cm]{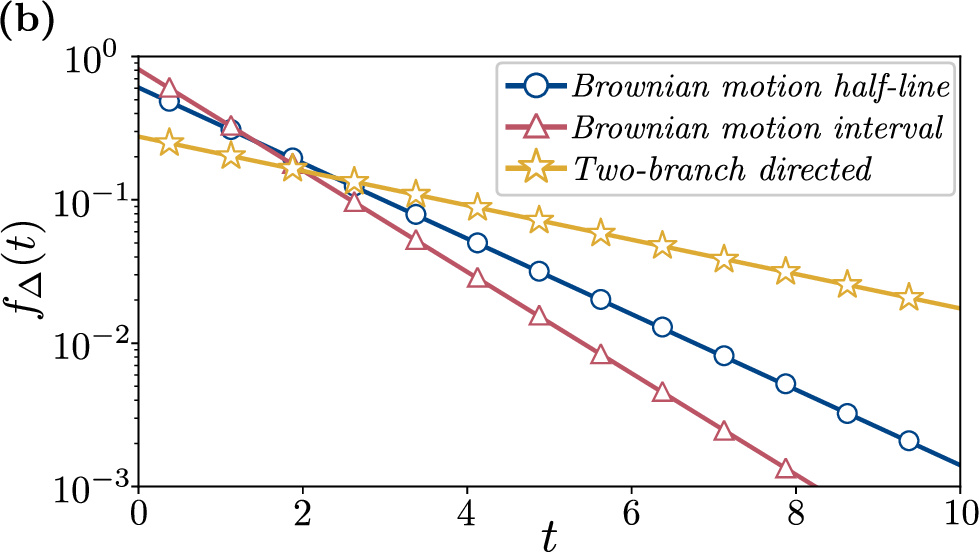}}
  \caption{Emergence of Markovianity in target inter-arrival times. (a) Convergence of the target arrival intensity function $a_{\lambda}(t)$ to steady-state. Solid curves denote the intensity function as in \cref{eq:IntensityFuncLambda}, markers denote Monte Carlo simulation data, and dashed lines denote the analytical steady-state as in \cref{eq:stat}. (b) Inter-arrival time density $f_{\Delta}(t)$. Solid lines denote the steady-state result in \cref{eq:IAT_density} and markers denote simulation data.}
  \label{fig:Fig6}
\end{figure}

In \cref{fig:Fig6_a} we illustrate this convergence to steady-state as well as the agreement of our theory to simulations. We make a few observations: For Brownian motion in the interval with uniformly distributed mortality times (example \labelcref{ex:b}), owing to the compact support of the mortality times, the probability flux into the target converges in finite time ($t\geq\mu/2$) to the exponential steady-state flux. In the case of directed search (example \labelcref{ex:c}), the probability flux into the target instantaneously converges to steady-state once time exceeds the minimum needed for an agent to find the target ($t\geq L/v$).

The limiting behavior in \cref{eq:stat} reveals a powerful fact about NCSS with homogeneous Poisson insertion: Target arrival times converge to a homogeneous Poisson process with rate $\lambda p$ and hence target inter-arrival times converge to an exponential distribution with rate $\lambda p$,
\begin{equation}
    \label{eq:IAT_density}
    f_{\Delta}(t) := \lim_{n\to\infty} \mathbb{P}(\Delta_n =t) =\lambda p e^{-\lambda p t}.
\end{equation}
Remarkably, the density in \cref{eq:IAT_density} only depends on the underlying stochastic search process insofar as those dynamics are encoded into the success probability $p$. We illustrate this in \cref{fig:Fig6_b} where we plot the steady-state target inter-arrival density against simulation data. 

%%%%%%%%%%%%%%%%%%%%%%%%%%%%%%%%%%%%%%%%%%%%%%%%%%%%%%%%%%%%%%%%%%%%%%%%%%%%%%%%%
%%%%%%%%%%%%%%%%%%%%%%%%%%%%%%%%%%%%%%%%%%%%%%%%%%%%%%%%%%%%%%%%%%%%%%%%%%%%%%%%%
%
%%%%%%%%%%%%%%%%%%%%%%%%%%%%%%%%%%%%%%%%%%%%%%%%%%%%%%%%%%%%%%%%%%%%%%%%%%%%%%%%%
%%%%%%%%%%%%%%%%%%%%%%%%%%%%%%%%%%%%%%%%%%%%%%%%%%%%%%%%%%%%%%%%%%%%%%%%%%%%%%%%%
\subsection{Fixing the average population recovers search-and-capture with additional structure}\label{subsec:4fix} Search-and-capture (SAC) is the process by which a stochastic agent serially delivers cargo from a fixed initial position to a target. Recent work characterizes the influence of stochastic resetting on SAC processes \cite{bressloff_queueing_2020, giral-barajas_resetting_2025}. By enforcing the non-conservative search process herein to have one agent on average, we recover the mean inter-arrival times of SAC processes with stochastic resetting in steady-state \cite{bressloff_queueing_2020}. The technical details are as follows.

In the interest of this comparison, suppose both the insertion and mortality distributions are exponential with rates $\lambda>0$ and $\mu>0$, respectively. The mean active search time therefore satisfies
\begin{equation} \label{eq:ES}
    \mathbb{E}[\mathcal{S}]=\int_0^\infty S_\mathcal{M}(t)S_\mathcal{T}(t)\,\textup{d}t = \frac{1-p}{\mu}.
\end{equation}
Enforcing that there be one agent on average requires carefully relating the insertion and mortality rates. Palm's theorem \cite{palm_intensity_1943} guarantees that, provided $\mathbb{E}[S]<\infty$,\footnote{In \cref{app:Mean} we show that the condition for Palm's theorem, $\mathbb{E}[S]<\infty$ with $\mathcal{S}=\min\{\mathcal{T},\mathcal{M}\}$, can be fulfilled even if $\mathbb{E}[\mathcal{M}]=\mathbb{E}[\mathcal{T}]=\infty$.} the limiting steady-state number of agents is Poisson distributed with parameter $\lambda\mathbb{E}[S]$; see \cref{app:queues} for more details. Maintaining one agent on average in steady-state therefore requires that
\begin{equation} \label{eq:lam}
    \lambda = \mathbb{E}[\mathcal{S}]^{-1} = \frac{\mu}{1-p}.
\end{equation}

\begin{figure}[b!]
  \centering
\includegraphics[width=0.585\textwidth]{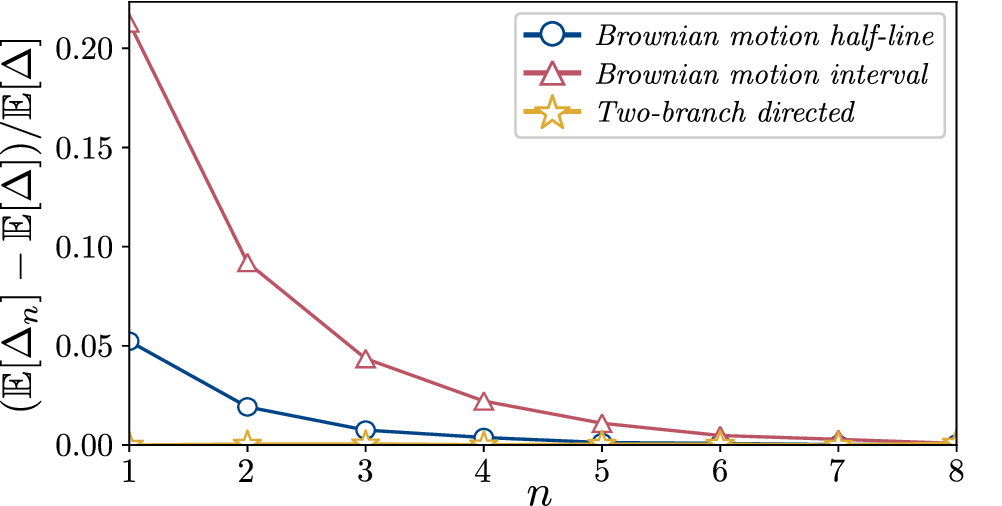}
  \caption{Convergence of the $\textup{n}$th mean target inter-arrival time of non-conservative stochastic search $\mathbb{E}[\Delta_n]$ to the SAC mean target inter-arrival time $\mathbb{E}[\Delta]$ as in \cref{eq:meanDelta}. The insertion rate, given by \cref{eq:lam}, is chosen such that there is one agent in $\Omega$ on average.}
  \label{fig:Fig7}
\end{figure}

This criterion is notably distinct from simply setting $\lambda=\mu$ since agents depart $\Omega$ not only through mortality but also through the target. Combining \cref{eq:IAT_density} with the criterion in \cref{eq:lam} yields exponentially distributed target inter-arrival times with mean 
\begin{equation}
    \label{eq:meanDelta}
    \mathbb{E}[\Delta]=\frac{1-p}{\mu p}.
\end{equation}
In \cref{fig:Fig7} we illustrate the rapid convergence of the running mean target inter-arrival time to the stationary value. We highlight that the expression in \cref{eq:meanDelta} is exactly the mean FPT of a single-agent SAC process with rate-$\mu$ stochastic resetting, but non-conservation provides even more structure. That is, while SAC processes often have general non-Markovian inter-arrival time distributions, non-conservative processes with a single agent on average recover the exponential mean-field behavior and gain distributional Markovianity in the target inter-arrival times.

%%%%%%%%%%%%%%%%%%%%%%%%%%%%%%%%%%%%%%%%%%%%%%%%%%%%%%%%%%%%%%%%%%%%%%%%%%%%%%%%%
%%%%%%%%%%%%%%%%%%%%%%%%%%%%%%%%%%%%%%%%%%%%%%%%%%%%%%%%%%%%%%%%%%%%%%%%%%%%%%%%%
%
%%%%%%%%%%%%%%%%%%%%%%%%%%%%%%%%%%%%%%%%%%%%%%%%%%%%%%%%%%%%%%%%%%%%%%%%%%%%%%%%%
%%%%%%%%%%%%%%%%%%%%%%%%%%%%%%%%%%%%%%%%%%%%%%%%%%%%%%%%%%%%%%%%%%%%%%%%%%%%%%%%%
\subsection{Monte Carlo simulations}\label{subsec:5MonteCarlo} We briefly detail how simulations were performed to produce the figures in this section. The simulation parameters used throughout are $\mu = D = L = 1$, $x_0=0.5$, $\rho_0=0.75$, and $v=1$. Figure-specific details are as follows. In \cref{fig:Fig5} we plot the explicit survival function using the quadrature approximation for example \labelcref{ex:a} and the closed form expressions of $\mathbb{A}(t)$ for examples \labelcref{ex:b,ex:c}. We simulate $N=10^6$ sample paths for each example, store the target FPTs and plot the empirical survival function,
\begin{equation}\label{eq:emp_surv}
    \hat{S}_N(t) = \frac{1}{N}\sum_{n=1}^N\mathbbm{1}_{\{\Delta_n\leq t\}}.
\end{equation}

In \cref{fig:Fig6_a} we use the values of $p$ from \cref{eq:exact_p} to produce the steady-state dashed lines. The time-dependent target fluxes (solid curves) are produced via quadrature for examples \labelcref{ex:a,ex:b} and using the explicit expression in example \labelcref{ex:c}. We use $N=10^6$ Monte Carlo approximations to simulate the time-dependent target flux. In \cref{fig:Fig6_b} we plot the explicit exponential density function with $p$ as in \cref{eq:exact_p}. Overlaid are $N=10^6$ Monte Carlo approximations of the density.

In \cref{fig:Fig7} we plot the relative distance between $\mathbb{E}[\Delta_n]$ for NCSS with $\lambda=\mu/(1-p)$ and $\mathbb{E}[\Delta]$ for SAC with rate-$\mu$ exponentially distributed stochastic resetting. As there is no explicit expression for $\mathbb{E}[\Delta_n]$ we rely on Monte Carlo simulations to sample the mean in each case using a sample size of $N=5\times10^6$.

\section{Discussion} \label{sec:discussion}
In this work we used multiple mathematical frameworks to derive and analyze the time-dependent and steady-state behavior of stochastic search processes with a fluctuating number of agents, referred to as non-conservative stochastic search (NCSS). The most general and streamlined approach involves constructing a map between NCSS and queueing theory, which enables a complete description of the time-dependent number of agents and target arrivals.

Assuming nonhomogeneous Poisson insertions of agents into the domain, the search process is exactly described by an $M_t/G/\infty$ queue. Utilizing results specific to this queue type, we derived an exact description of agent arrivals to the target and found it to follow a nonhomogeneous Poisson law. This result subsequently allowed for computing the time-dependent probability flux into the target and its convergence to steady-state. These results collectively reveal a universal feature of NCSS: steady-state inter-arrival times to the target are exponentially distributed, regardless of the underlying stochastic search dynamics. Finally, we showed that fixing the mean steady-state number of agents to be one recovers the mean inter-arrival time of single-agent search-and-capture processes while gaining Markovianity in the law.

One future application of the results presented here is target resource accumulation under different consumption or utilization protocols. Following previous work on sequential search-and-capture \cite{bressloff_search-and-capture_2019, bressloff_directional_2021, bressloff_first-passage_2021, bressloff_queueing_2020, bressloff_queuing_2021, giral-barajas_stochastic_2025, giral-barajas_resetting_2025}, the target itself can be treated as a downstream queue. The target arrival statistics then represent customers entering a queue and the consumption protocol represents customer servicing. The latter is typically modeled in terms of Poisson process involving $c$ servers where $c$ may be infinite. Resource accumulation under sequential search-and-capture typically maps to a $G/M/c$ queue. An immediate consequence of our work is that resource accumulation for NCSS processes is universally described by an $M/M/c$ queue. This greatly simplifies the analysis, which is particularly meaningful when considering splitting probabilities and resource accumulation across multiple targets as well as blow-up and vacancy times at target sites. Another interesting extension would be to reformulate agents as responsible for both resource delivery and service. That is, rather than assuming the server number at a target to be independent of the agents, one would like to characterize resource accumulation when agents intermittently process resources between successive delivery excursions.

\section*{Acknowledgments}
JGB was supported by a Roth PhD scholarship funded by the Department of Mathematics at Imperial College London. SL was supported by the U.S. National Science Foundation grant DMS-2503350. This research was also supported in part by grants from the NSF (DMS-2235451) and Simons Foundation (MPS-NITMB-00005320) to the NSF-Simons National Institute for Theory and Mathematics in Biology (NITMB).

%%%%%%%%%%%%%%%%%%%%%%%%%%%%%%%%%%%%%%%%%%%%%%%%%%%%%%%%%%%%%%%%%%%%%%%%%%%%%%%%%
%%%%%%%%%%%%%%%%%%%%%%%%%%%%%%%%%%%%%%%%%%%%%%%%%%%%%%%%%%%%%%%%%%%%%%%%%%%%%%%%%
%
%%%%%%%%%%%%%%%%%%%%%%%%%%%%%%%%%%%%%%%%%%%%%%%%%%%%%%%%%%%%%%%%%%%%%%%%%%%%%%%%%
%%%%%%%%%%%%%%%%%%%%%%%%%%%%%%%%%%%%%%%%%%%%%%%%%%%%%%%%%%%%%%%%%%%%%%%%%%%%%%%%%

\appendix
\crefalias{section}{appendix}
\crefalias{subsection}{appendix}
\section{Notation and definitions}\label{app:notation} \Cref{tab:notation} summarizes the important definition and the choice of notation. 

\begin{table}[htbp]
\footnotesize
  \caption{Non-conservative search processes' notation and definitions}\label{tab:notation}
\begin{center}
  \begin{tabular}{cl} \hline\hline
  \multicolumn{2}{c}{\textbf{Non-conservative search stochastic processes (\textit{Capital Latin} letters)}}\\
  \hline
  $I(t)$ & Number of inserted agents into $\Omega$ up to time $t$.\\
  $Q(t)$ & Number of agents in $\Omega$ at time $t$.\\
  $D(t)$ & Number of departures from $\Omega$ up to time $t$ (arrivals to target and mortality events).\\
  $A(t)$ & Number of arrivals to target, $\Omega_0$, up to time $t$.\\
  $M(t)$ & Number of mortality events up to time $t$.\\
  \hline
  \multicolumn{2}{c}{\textbf{Mean functions (\textit{Blackboard bold} letters)}}\\
  \hline
  $\mathbb{I}(t)$ & Expected number of inserted agents into $\Omega$ up to time $t$.\\
  $\mathbb{Q}(t)$ & Expected number of agents in $\Omega$ at time $t$.\\
  $\mathbb{D}(t)$ & Expected number of departures from $\Omega$ up to time $t$.\\
  $\mathbb{A}(t)$ & Expected number of arrivals to target, $\Omega_0$, up to time $t$.\\
  $\mathbb{M}(t)$ & Expected number of mortality events up to time $t$.\\
  \hline
  \multicolumn{2}{c}{\textbf{Non-conservative search service-time r.v.s (\textit{Calligraphic} letters)}}\\
  \hline
  $\mathcal{M}$ & Lifetime of a mortal agent.\\
  $\mathcal{T}$ & FPT without mortality.\\
  $\mathcal{S}$ & Time spent in $\Omega$ by an agent, \textit{i.e.}~service time. Defined as $\min\{\mathcal{T},\mathcal{M}\}$ .\\
  \hline\hline
  \end{tabular}
\end{center}
\end{table}

%%%%%%%%%%%%%%%%%%%%%%%%%%%%%%%%%%%%%%%%%%%%%%%%%%%%%%%%%%%%%%%%%%%%%%%%%%%%%%%%%
%%%%%%%%%%%%%%%%%%%%%%%%%%%%%%%%%%%%%%%%%%%%%%%%%%%%%%%%%%%%%%%%%%%%%%%%%%%%%%%%%
%
%%%%%%%%%%%%%%%%%%%%%%%%%%%%%%%%%%%%%%%%%%%%%%%%%%%%%%%%%%%%%%%%%%%%%%%%%%%%%%%%%
%%%%%%%%%%%%%%%%%%%%%%%%%%%%%%%%%%%%%%%%%%%%%%%%%%%%%%%%%%%%%%%%%%%%%%%%%%%%%%%%%

\section{Proof of \texorpdfstring{{\boldmath$p>0$}}{p>0}}\label{app:p_proof} Here we show a set of sufficient conditions ensuring the probability that a single agent finds the target before dying is positive, \textit{i.e.}~$p>0$. We need that the FPT without mortality is not always greater than the maximum lifetime. This is, the support of $\mathcal{M}$ needs to intersect the support of $\mathcal{T}$, which can be formally written as
\begin{equation}\label{eq:sup-inf-cond}
    \essinf(\mathcal{T})<\esssup(\mathcal{M}),
\end{equation}
where $\esssup(X)$ and $\essinf(X)$ are the \textit{essential supremum} and \textit{essential infimum} of a r.v.~$X$, defined by
\begin{subequations}
\begin{eqnarray}
    \esssup(X) & := & \inf\{x\in\mathbb{R}\mid \mathbb{P}(X>x)=0\},\label{eq:esssup-def}\\
    \essinf(X) & := &\sup\{x\in\mathbb{R}\mid \mathbb{P}(X\geq x)=1\}.\label{eq:essinf-def}
\end{eqnarray}
\end{subequations}

We show that, given the assumption made in the setting, \cref{eq:sup-inf-cond} is a sufficient condition to achieve $p>0$. First, as $\mathcal{T}$ is strictly positive, we know that $0<\essinf(\mathcal{T})$. As $\essinf(\mathcal{T})<\esssup(\mathcal{M})$, there exists $\varepsilon>0$ such that
\begin{equation}\label{eq:sup-inf-epsilon}
    \essinf(\mathcal{T})<\varepsilon<\esssup(\mathcal{M}).
\end{equation}
Since $\varepsilon < \esssup(\mathcal{M})$, by definition of the essential supremum,
\begin{equation}\label{eq:cond1}
    S_\mathcal{M}(\varepsilon)=\mathbb{P}(\mathcal{M}>\varepsilon)>0.
\end{equation}
Moreover, since $S_\mathcal{M}$ is right-continuous and non-increasing with $S_\mathcal{M}(0)=1$, $S_\mathcal{M}(t)\geq S_\mathcal{M}(\varepsilon)>0$ for all $t\in[0,\varepsilon)$. On the other hand, since $\varepsilon > \essinf(\mathcal{T})$, by definition of the essential infimum, 
\begin{equation}\label{eq:cond2}
    F_\mathcal{T}(\varepsilon)=\mathbb{P}(\mathcal{T}\leq\varepsilon)>0.
\end{equation}
This is, $F_\mathcal{T}$ assigns strictly positive mass to $(0,\varepsilon]$. Therefore, combining \cref{eq:cond1} and \cref{eq:cond2} with the definition of $p$ \cref{eq:pmu}, we have
\begin{equation}\label{eq:pmu_proof}
   p= \int_0^\infty S_\mathcal{M}(t)\,\textup{d}F_\mathcal{T}(t)\geq \int_0^\varepsilon S_\mathcal{M}(t)\,\textup{d}F_\mathcal{T}(t) \geq S_\mathcal{M}(\varepsilon)F_\mathcal{T}(\varepsilon)>0.
\end{equation}

\subsection{Examples fulfilling the necessary condition}\label{app:examples_p} It is important to note that $\essinf(\mathcal{T})$ depends on the stochastic dynamics, the geometry of $\Omega$, and the starting position, $\mathbf{x}_0$. Therefore, in general, the conditions to fulfill \cref{eq:sup-inf-cond} are specific to the search process. To finalize this section, we provide some general settings where the sufficient condition \cref{eq:sup-inf-cond} is fulfilled and $p>0$.

\paragraph{Unbounded mortality support} A trivial, yet profoundly useful, scenario is one in which the mortality r.v.~has unbounded support. This is a very weak assumption about $\mathcal{M}$, allowing it to adopt a wide range of distributions. In this setting, $\esssup(\mathcal{M)}=\infty$. As $\mathcal{T}$ is assumed to be not always infinite, $\essinf(\mathcal{T)}<\infty$. Therefore, if the lifetime $\mathcal{M}$ has unbounded support \cref{eq:sup-inf-cond} is automatically fulfilled, and $p>0$ is guaranteed regardless of the stochastic dynamics, the geometry of $\Omega$, or the starting position $\mathbf{x}_0$.

\paragraph{Search processes with infinite propagation speed} For stochastic processes with infinite propagation speed, such as Brownian motion and L\'evy flights, the transition density $p(\mathbf{x}, t|\mathbf{x}_0)$ is strictly positive for all $\mathbf{x}$ and all $t > 0$. Therefore $\text{ess inf}(\mathcal{T}) = 0$, regardless of the geometry of $\Omega$, the location of $\Omega_0$, and the starting position $\mathbf{x}_0$. As $\mathcal{M}$ is strictly positive, $\text{ess sup}(\mathcal{M}) > 0$. Consequently, \cref{eq:sup-inf-cond} is automatically fulfilled, and $p>0$ is guaranteed.

\paragraph{Search processes with finite speed} For stochastic processes with finite speed, such as run-and-tumble particles or directed search, we define $v^*$ as the maximum speed from $\mathbf{x}_0$ to $\Omega_0$. In this case,
\begin{equation}
    \text{ess inf}(\mathcal{T}) = \frac{\text{dist}(\mathbf{x}_0,\Omega_0)}{v^*},
\end{equation}
and $\mathcal{M}$ must fulfill 
\begin{equation}\label{eq:bounded_cond}
    \text{ess sup}(\mathcal{M}) > \frac{\text{dist}(\mathbf{x}_0,\Omega_0)}{v^*}.
\end{equation}
This is only relevant when $\mathcal{M}$ has a bounded domain $[a,b]$ with $0< a<b<\infty$. In this case \cref{eq:bounded_cond} is equivalent to $b>\text{dist}(\mathbf{x}_0,\Omega_0)/v^*$.

%%%%%%%%%%%%%%%%%%%%%%%%%%%%%%%%%%%%%%%%%%%%%%%%%%%%%%%%%%%%%%%%%%%%%%%%%%%%%%%%%
%%%%%%%%%%%%%%%%%%%%%%%%%%%%%%%%%%%%%%%%%%%%%%%%%%%%%%%%%%%%%%%%%%%%%%%%%%%%%%%%%
%
%%%%%%%%%%%%%%%%%%%%%%%%%%%%%%%%%%%%%%%%%%%%%%%%%%%%%%%%%%%%%%%%%%%%%%%%%%%%%%%%%
%%%%%%%%%%%%%%%%%%%%%%%%%%%%%%%%%%%%%%%%%%%%%%%%%%%%%%%%%%%%%%%%%%%%%%%%%%%%%%%%%

\section{Fundamentals of queues with Poisson insertions and an infinite number of servers}\label{app:queues}

\subsection{General setting and notation}
A queueing process tracks the number of agents present in a system that provides some type of service, as a function of time. Insertion and service times are generally random, so describing such a process requires specifying the insertion-time distribution, the service-time distribution, the number of servers, the queue discipline, and the system capacity \cite{allen_probability_1978}. Throughout this work we assume infinite capacity, \textit{i.e.}~the queue can hold an infinite number of agents, and a first-come, first-served discipline, so that a queueing system is fully determined by Kendall's short-hand notation $A/B/c$, where $A$ denotes the inter-insertion time distribution, $B$ the service-time distribution, and $c$ the number of servers \cite{kendall_stochastic_1953}. For example, $M/G/\infty$ denotes a queueing system with exponentially distributed (Markovian) inter-insertion times, service times with a general (non-Markovian) distribution and an infinite number of servers.

We further assume an infinite number of servers ($c=\infty$). Therefore, every agent accesses a server and begins service immediately upon insertion, with no waiting time. The time an agent spends in the system is thus simply its service time, denoted $\mathcal{S}$ throughout this work, after which it departs. Writing $I(t)$ and $D(t)$ for the number of insertions and departures in $(0,t]$, the number of agents still receiving service at time $t$, corresponding to the queue length, is
\begin{equation}
    Q(t) = I(t) - D(t), \qquad t>0.
\end{equation}

As shown in \cref{sec:queue1}, the number of active agents in $\Omega$ for a non-conservative search process maps onto exactly this type of infinite-server queue: insertions follow a nonhomogeneous Poisson process $\{I(t):t\geq0\}$ with intensity function $\lambda(t)$, and the service-time distribution is general (non-Markovian). In Kendall's notation, this corresponds to an $M_t/G/\infty$ queueing system. If the insertion process is simplified to a homogeneous Poisson process of rate $\lambda$, \textit{i.e.}~$\lambda(t)\equiv\lambda$, the inter-insertion time distribution reduces to the exponential distribution, and the system instead corresponds to an $M/G/\infty$ queue. We present the relevant results on the stationary number of agents and on the departure process for the $M_t/G/\infty$ and $M/G/\infty$ queues in turn, in the following two subsections.

%%%%%%%%%%%%%%%%%%%%%%%%%%%%%%%%%%%%%%%%%%%%
\subsection{Nonhomogeneous Poisson insertions (\texorpdfstring{{\boldmath$M_t/G/\infty$}}{Mt/G/∞})} As before, assume that the insertion process is a nonhomogeneous Poisson process with rate $\lambda(t)$, so that the corresponding queueing system is $M_t/G/\infty$. This system has been extensively studied \cite{eick_physics_1993,foley1982non,foley_stationary_1986}; in particular, for each $t>0$ the queue length $Q(t)$ is known to be Poisson distributed, with mean
\begin{equation}
    \label{eq:MtQueueLength}
    Q(t)\sim\text{Poisson}\left(\int_0^tS_{\mathcal{S}}(t-t')\,\lambda(t')\,\textup{d}t'\right),
\end{equation}
where $S_{\mathcal{S}}(t)$ denotes the service-time survival function.

The central result we rely on for this work concerns the departure process. This result states that the departure process $\{D(t):t\geq0\}$ of an $M_t/G/\infty$ queue is itself a nonhomogeneous Poisson process, with expected number of departures in $(0,t]$ given by
\begin{equation}
    \label{eq:departureMean}
    \mathbb{D}(t):=\mathbb{E}[D(t)]=\int_0^t F_{\mathcal{S}}(t-t')\,\lambda(t')\,\textup{d}t',
\end{equation}
where $F_\mathcal{S}(t) = 1- S_{\mathcal S}(t)$ denotes the service-time distribution function \cite{foley1982non}. This result is a stronger statement than \cref{eq:MtQueueLength}. Rather than describing the distribution of the number of agents present at a fixed time, this result characterizes the full law of the departure process, including independent increments and the explicit time-mean $\mathbb{D}(t)$. It is this process-level Poisson structure for $D(t)$ that we exploit to derive our main results. For a proof of the results presented in \cref{eq:MtQueueLength,eq:departureMean} see \cite[Theorem 1]{eick_physics_1993}.

%%%%%%%%%%%%%%%%%%%%%%%%%%%%%%%%%%%%%%%%%%%%
\subsection{Homogeneous Poisson insertions (\texorpdfstring{{\boldmath$M/G/\infty$}}{M/G/∞})} The $M/G/\infty$ queue is the special case of the system above obtained by setting $\lambda(t)\equiv\lambda$, so that insertions follow a homogeneous Poisson process and inter-insertion times are exponentially distributed. It is in fact the older of the two systems, and its properties were characterized by Palm \cite{palm_intensity_1943} and Takács \cite{takacs_introduction_1962, takacs_coincidence_1958} well before the nonhomogeneous generalization was established. Since then, the $M/G/\infty$ queue has been extensively studied; see also \cite{mirasol_output_1963,brown_results_1969,carrillo_extensions_1991}. In particular, the special case of \cref{eq:MtQueueLength} with a constant rate recovers the classical result that, for each $t>0$, the queue length is Poisson distributed with mean
\begin{equation}
    \label{eq:PoissonMeanTakacs}
    Q(t)\sim\text{Poisson}\left(\lambda\int_0^t S_{\mathcal{S}}(t')\,\textup{d}t'\right).
\end{equation}

Unlike the nonhomogeneous case, it is easy to move from the time-dependent queue length to the steady-state process. Palm's theorem \cite{palm_intensity_1943} shows that, provided $\bar{\mathcal{S}}:=\mathbb{E}[\mathcal{S}]<\infty$, the limiting probability $P_k^*:=\lim_{t\to\infty}\mathbb{P}(Q(t)=k)$ exists for every $k\geq0$ and is Poisson distributed with parameter $\lambda\bar{\mathcal{S}}$,
\begin{equation}
    \label{eq:PoissonSSProbabilityTakacs}
    P_k^* = e^{-\lambda\bar{\mathcal{S}}}\frac{\left[\lambda\bar{\mathcal{S}}\right]^k}{k!}.
\end{equation}
See \cite[Chapter 3.2]{takacs_introduction_1962} for a proof.

It has been showed that the departure process $\{D(t):t\geq0\}$ of the $M/G/\infty$ queue has the same Poisson structure established above for the nonhomogeneous case and is itself a nonhomogeneous Poisson process \cite{mirasol_output_1963}, with
\begin{equation}
    \label{eq:GMC_departure}
    \mathbb{D}(t):=\mathbb{E}[D(t)]=\lambda\int_0^t F_{\mathcal{S}}(t')\,\textup{d}t',
\end{equation}
a proof of which can be found in \cite[Example 5.25]{ross_introduction_2010}. As before, this is a statement about the full law of the departure process rather than its one-time marginal; together with \cref{eq:PoissonMeanTakacs,eq:PoissonSSProbabilityTakacs}, it completes the description of the $M/G/\infty$ queue in both the transient and steady-state regimes.
%%%%%%%%%%%%%%%%%%%%%%%%%%%%%%%%%%%%%%%%%%%%%%%%%%%%%%%%%%%%%%%%%%%%%%%%%%%%%%%%%
%%%%%%%%%%%%%%%%%%%%%%%%%%%%%%%%%%%%%%%%%%%%%%%%%%%%%%%%%%%%%%%%%%%%%%%%%%%%%%%%%
%
%%%%%%%%%%%%%%%%%%%%%%%%%%%%%%%%%%%%%%%%%%%%%%%%%%%%%%%%%%%%%%%%%%%%%%%%%%%%%%%%%
%%%%%%%%%%%%%%%%%%%%%%%%%%%%%%%%%%%%%%%%%%%%%%%%%%%%%%%%%%%%%%%%%%%%%%%%%%%%%%%%%

\section{Finite active search-time mean minimal assumptions}\label{app:Mean} Let $Y,Z$ be two independent, non-negative random variables. Define $X:=\min\{Y,Z\}$. In this appendix we explore what are the minimal assumptions for $Y$ and $Z$ ensuring that $\mathbb{E}[X]<\infty$. First, it is clear that, as $X\leq Y,Z$, $\mathbb{E}[X]<\mathbb{E}[Y]$ and $\mathbb{E}[X]<\mathbb{E}[Z]$. Therefore, the assumption that $\mathbb{E}[Y]<\infty$ or $\mathbb{E}[Z]<\infty$ is sufficient to ensure that $\mathbb{E}[X]<\infty$. However, this is not a necessary condition. A simple counterexample can be constructed by considering two independent Type I Pareto random variables with shape and scale parameters equal to one.

If $Y$ is a Type I Pareto random variable with scale parameter $x_{min}$ and shape parameter $\alpha$, denoted by $Y\sim\text{Pa}(x_{min},\alpha)$, its survival probability is given by
\begin{equation}
    \label{eq:Pareto}
    S_Y(x)=\begin{cases}
        \left(\frac{x_{min}}{x}\right)^\alpha\quad&x\geq x_{min},\\
        1\quad&x<x_{min}.
    \end{cases}
\end{equation}
It is known that if $Y\sim\text{Pa}(x_{min},\alpha)$ then
\begin{equation}
    \label{eq:ParetoMean}
    \mathbb{E}[Y]=\begin{cases}
        \infty\quad&\alpha\leq1,\\
        \frac{\alpha x_{min}}{\alpha-1}\quad&\alpha>1.
    \end{cases}
\end{equation}
Now, we assume that $Y,Z\sim\text{Pa}(1,1)$ and $Y$ is independent of $Z$. Following \cref{eq:ParetoMean} we have that $\mathbb{E}[Y]=\mathbb{E}[Z]=\infty$. On the other hand, since $Y$ and $Z$ are independent, we have that
\begin{equation}
    \label{eq:SurvivalX}
    S_X(x)=S_Z(x)S_Y(x)=\begin{cases}
        \left(\frac{1}{x}\right)^2\quad&x\geq 1,\\
        1\quad&x<1.
    \end{cases}
\end{equation}
With this explicit expression of the survival probability, we directly compute $\mathbb{E}[X]$ as
\begin{equation}
    \mathbb{E}[X] = \int_1^\infty x^{-2}\,\textup{d}x=1.
\end{equation}
This finalizes the counterexample in which $\mathbb{E}[Y]=\mathbb{E}[Z]=\infty$, but $\mathbb{E}[X]<\infty$. 

In the general case in which $Y$ and $Z$ are two non-negative and independent random variables we still have the following expression for the mean of X
\begin{equation}
    \label{eq:Xmean}
    \mathbb{E}[X] = \int_0^\infty S_Y(x)S_Z(x)\,\textup{d}x.
\end{equation}
Therefore the necessary and sufficient condition on $Y$ and $Z$ for $X$ to have finite mean is
\begin{equation}
    \label{eq:FiniteMeanCondition}
    \int_0^\infty S_Y(x)S_Z(x)\,\textup{d}x<\infty.
\end{equation}

\bibliographystyle{unsrt}
\bibliography{references}
\end{document}

%% file: ex_shared.tex
\usepackage{lipsum}
\usepackage{amsfonts}
\usepackage{graphicx}
\usepackage{epstopdf}
\usepackage{algorithmic}
\usepackage[shortlabels]{enumitem}
\usepackage{bbm}
\usepackage[bottom]{footmisc}
\ifpdf
  \DeclareGraphicsExtensions{.eps,.pdf,.png,.jpg}
\else
  \DeclareGraphicsExtensions{.eps}
\fi

\newsiamremark{remark}{Remark}
\newsiamremark{hypothesis}{Hypothesis}
\crefname{hypothesis}{Hypothesis}{Hypotheses}
\newsiamthm{claim}{Claim}
\newsiamremark{fact}{Fact}
\crefname{fact}{Fact}{Facts}

\headers{Universality in arrival statistics of non-conservative search}{J.~Giral-Barajas, S.~Linn, and P.~C.~Bressloff}

\title{Universality in the target arrival statistics of non-conservative search processes\thanks{Submitted to the editors \today.
\funding{JGB was supported by a Roth PhD scholarship funded by the Department of Mathematics at Imperial College London. SL was supported by the U.S.\ National Science Foundation grant DMS-2503350.}}}

\author{Jos\'e Giral-Barajas\thanks{Department of Mathematics, Imperial College London, London SW7 2AZ, UK (\email{j.giral-barajas24@imperial.ac.uk}, \email{s.linn@imperial.ac.uk}, \email{p.bressloff@imperial.ac.uk}).}
\and Samantha Linn\footnotemark[2]
\and Paul C. Bressloff\footnotemark[2]}

\usepackage{amsopn}

\DeclareMathOperator{\esssup}{ess\,sup}
\DeclareMathOperator{\essinf}{ess\,inf}

\usepackage{amsmath}
\usepackage{amssymb}
\usepackage{amsfonts}
\usepackage{graphicx,epstopdf}
\usepackage[caption=false]{subfig}